# Local sensing of absolute refractive index during protein-binding using microlasers with spectral encoding

*Soraya Caixeiro[1,2]\*, Casper Kunstmann-Olsen[3], Marcel Schubert[1,2], Joseph Hill[1], Isla R. M. Barnard[1], Matthew D. Simmons[3], Steven Johnson[3,4]\*, Malte C. Gather [1,2]\**

[1] SUPA, School of Physics & Astronomy, University of St. Andrews, North Haugh, St Andrews, KY16 9SS, UK.
[2] Humboldt Centre for Nano- and Biophotonics, Department of Chemistry, University of Cologne, Greinstr. 4-6, 50939 Cologne, Germany.
[3] Department of Electronic Engineering, University of York, Heslington, York, YO10 5DD, UK
[4] York Biomedical Research Institute.
\* soraya.caixeiro@uni-koeln.de; steven.johnson@york.ac.uk; mcg6@st-andrews.ac.uk

Abstract:
Multiplexed, specific and sensitive detection of antigens is critical for the rapid and accurate diagnosis of disease and the informed development of personalized treatment plans. Here, we show that polymer microsphere lasers can be used as photonic sensors to monitor and quantify direct surface binding of biomolecules via changes in the refractive index. The unique spectral signature of each individual laser can be used to find their size and effective refractive index which adds a new encoding dimension when compared to conventional fluorescent beads. We utilize antibody-functionalized microlasers to selectively detect protein binding. Different stages of the multilayer surface modification can be resolved, and protein binding is demonstrated for two different proteins, IgG and CRP. Moreover, by continuously monitoring single lasers, we demonstrate the possibility of real-time monitoring of binding dynamics between antigens in solution phase and the immobilized antibodies. For multiplexed detection, the microlasers are employed in a flow cytometer configuration, with fast spectral detection and identification of microlasers with and without antigen binding. We envision that by combining microlasers with well-established surface modification chemistries and flow geometries, the multiplexing ability of microbead immunoassays can be strongly increased while also opening avenues for single cell profiling within heterogenous cell populations.

# 1    Introduction

High throughput, multiplexed screening of biomarkers is driving numerous advances in the fields of cellular biology, immunology, environmental testing, and drug research. One approach to this is flow cytometry using polymer microspheres[1–5] or liquid droplets.[6–8] For example multiplexed detection is often achieved by using fluorescent microspheres with different emission colors, each encoded with a distinct capture antibody. A secondary fluorescently labelled antibody is then used to confirm the presence of the target molecule. The number of spectral codes that can be separated is generally limited by the spectral width of the emission spectra associated with each label; for organic dyes, the spectra are typically > 30 nm wide and using different organic dyes one can readily cover most of the visible and a small part of the near infrared spectrum, i.e., approximately 420 to 800 nm. Strategies to enlarge the multiplexing capacity include microspheres containing several different dyes and different dye ratios,[9,10] and using emitters with narrower spectra such as quantum dots[11,12] and lanthanides,[13,14] where the latter has allowed for more than 1000 distinguishable spectral outputs. Nevertheless, a fluorescently tagged antibody would be required for antigen detection, which adds complexity and cost to the experiment and precludes real-time monitoring of antigen binding kinetics.

Whispering gallery mode (WGM) resonators have been used for a range of biochemical sensors[15–19] and in commercial immunoassays [20] which are summarized in many excellent reviews.[21,22] However, characteristically this required large resonators (10s μm to mm in diameter), complex coupling geometries, continuous monitoring with tunable external light sources, and relied on relative wavelength shifts.

WGM microlasers generally exhibit narrower linewidths when operated above lasing threshold in comparison to conventional fluorescence and emission from a passive WGM cavity. This spectrally well-defined emission can facilitate a high degree of multiplexing, while the size and bright emission of WGM lasers allows for non-diffusion limited reactions, flow cytometry integration and remote detection. Biointegrated WGM microlasers have recently been applied for label-free tagging and tracking of cells and intracellular sensing.[23–28] Refractive index sensing with WGM lasers has been explored for a number of applications, such as cardiac contractility sensing[25] as well as protein[29] and single virus detection.[30]

In this work, we propose the use of surface-modified polystyrene (PS) WGM microlasers for the detection of biomarkers in aqueous environments via an antibody-antigen sensing platform. The distinct spectral profile is accurately described via an asymptotic expansion,[31] yielding both size and external refractive index as independent parameters. Microlasers are used as absolute refractive index reporters at different stages of the surface modification and to monitor the antigen binding dynamics through the real-time change of refractive index of the microlaser environment in the presence of the desired target protein. We show proof of concept implementation of the technique in a flow cytometer-type instrument, allowing us to distinguish different stages of surface modification with high throughput. We envision that by combining lasing microspheres with surface modifications and flow geometries already employed by the scientific community, it will be possible to extend the multiplexing capability beyond the level achievable with currently used methods.

## 2     Results and Discussion

### 2.1     Microsphere Functionalization

The WGM microlasers used in this study are made from commercially available PS microspheres and are approximately 11 µm in diameter. Optical gain is provided by dye molecules embedded into the microsphere via swelling and encapsulation,[32] with an absorption maximum at 441 nm. The microspheres have carboxyl groups on their surface, which we used for the covalent coupling of antibodies and antigens of interest. To achieve antibody coupling, the carboxylated PS microlasers were added to a carbodiimide and an ester precursor (NHS/EDC) to first form an amino-reactive NHS-ester at the surface as depicted in Figure 1a. The NHS-functionalized microlaser surface allows the binding of free amines on the antibody surface enabling covalent attachment to the microlaser. The microlaser is afterwards incubated with casein to block exposed regions of the microsphere surface and to help inhibit non-specific binding of antigen to the PS surface. As a control, instead of binding the antibody of interest, the surface is completely blocked with casein, thus preventing specific binding of the antigen of interest (referred to as blocked microlasers). Two different antigens are used in this work, C-reactive protein (CRP) and Immunoglobulin G (IgG). CRP is an important biomarker usually indicative of inflammation and/or infection[33], while IgG is a common antibody found in human blood, which has many uses such as binding and neutralization of toxins[34]. (The functionalization protocol and its optimization are detailed in the Experimental section and the Supplementary Information Figures S1-S5).

To verify the successful functionalization, a fluorescent label (NHS-ester-tagged Alexa-568) is added to the antigen and incubated with the corresponding antibody labelled microlasers. The maximum emission of the fluorescently labelled antigen ($\lambda = 603$ nm) is sufficiently isolated from the emission maxima of the dye inside the microspheres ($\lambda \sim 490$ nm and $\lambda \sim 515$ nm) to allow clear spectral separation. Figure 1b and 1c show the green and red channel fluorescence, respectively. The ring-shaped red fluorescence indicates good surface coverage of the green-fluorescing microspheres with the red-emitting antigen.

To evaluate the specificity of the functionalization, samples of antibody functionalized microspheres and blocked microspheres are processed according to the protocol described in the Experimental section. The average intensity in the red epi-fluorescence images was then compared (Figure 1d). The blocked microlasers show a minor increase of red fluorescence when incubated with fluorescent antigens, which we attribute to a small degree of non-specific binding. By contrast, when incubating with antibody-functionalized microlasers, there is a strong increase in the average fluorescence intensity. These results correlate well with experiments conducted using a quartz crystal microbalance setup (Supplementary Information Fig. S3 and S5) and confirm the successful binding of antibodies to the microlaser surface.

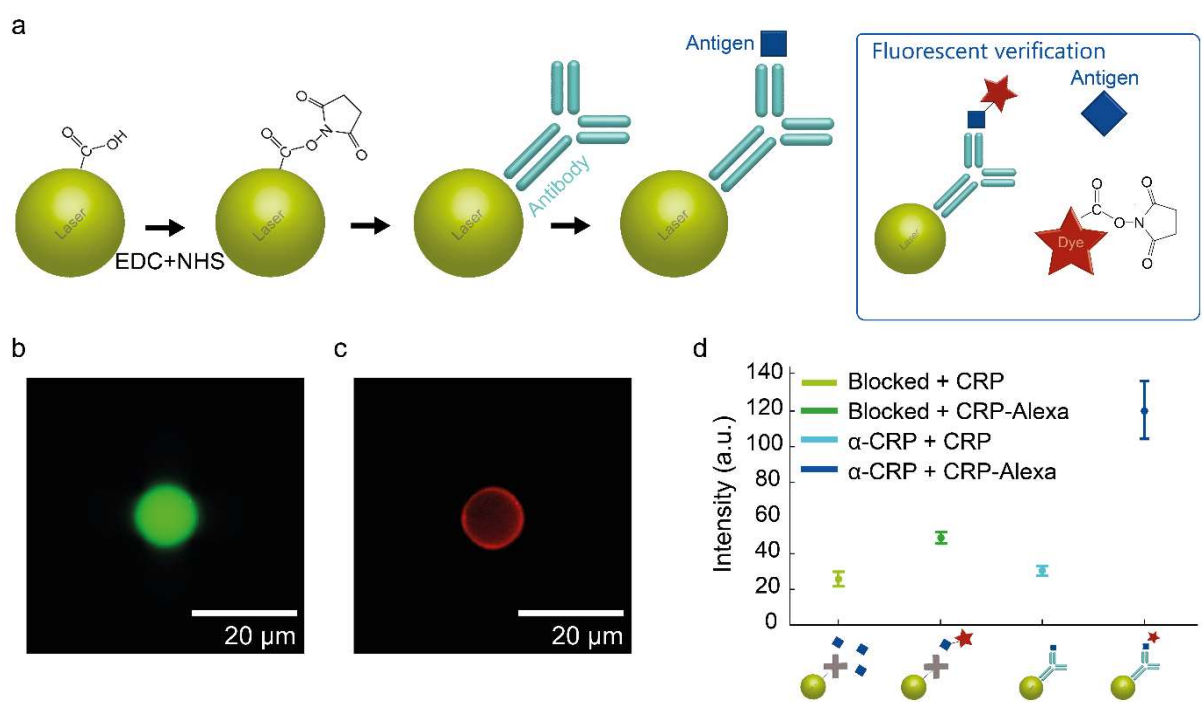

**Figure 1.** Functionalization of carboxylated polystyrene microlasers. (a) Schematic of the microlaser functionalization. The inset shows the fluorescent labelling scheme used to validate antigen binding to the antibody-functionalized microlaser surface. An NHS functionalized fluorescent tag (symbolized by red star) is used to bind to the antigen (blue square). (b, c) Epi-fluorescence microscopy images of antibody-functionalized microlasers, taken in (b) the green bulk fluorescence and (c) the red Alexa-568 fluorescence channel. (d) Average red fluorescence intensity of a series of blocked microlasers incubated with the untagged antigen (light green) and fluorescently tagged antigen CRP (dark green), as well as antibody-functionalized microlasers incubated with untagged antigen (light blue) and fluorescently tagged antigen (dark blue). $N \geq 10$ for each group of microlasers.

## 2.2   Laser Characterization

Microspheres were optically characterized using a custom-built microscopy/spectroscopy setup equipped with a nanosecond pulsed laser system tuned to 440 nm (see Experimental section). Our microlasers showed an average threshold fluence for lasing of 130 µJ cm$^{-2}$ ± 40 µJ cm$^{-2}$ ($N = 5$, corresponding to an absolute threshold pulse energy of 5 nJ), similar to other organic WGM microlasers reported in the literature.[24,35,36] Figure 2a depicts a representative threshold input-output curve fitted with a rate equation model.[37] Figure 2b shows the typical spectral evolution of the emission from the microspheres for selected pump powers. Below threshold, the spectrum is initially dominated by a background of fluorescence, with resonances emerging from the background when approaching the laser threshold. Above threshold, lasing occurs for some of these resonances. At higher pulse energies lasing from these resonances dominates the spectrum, with an intensity typically 2-3 orders of magnitude above the background fluorescence.

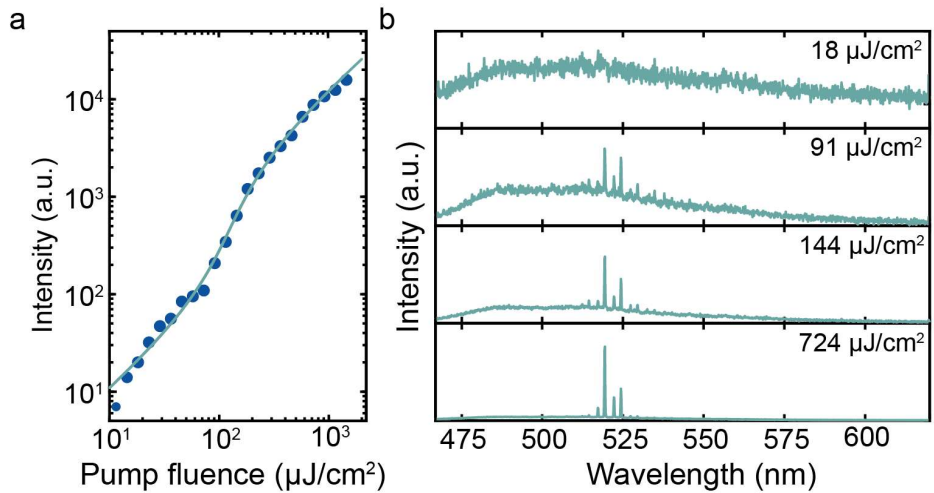

**Figure 2.** Characteristics of a carboxylated microsphere laser in PBS solution. (a) Output intensity of a microlaser as a function of incident pump fluence. The green line represents a fit to the rate equation model, with the fitted parameters indicating a threshold pump fluence of 125 µJ cm$^{-2}$. (b) WGM spectrum of the microlaser below (18 µJ cm$^{-2}$ and 91 µJ cm$^{-2}$), around (144 µJ cm$^{-2}$), and above (724 µJ cm$^{-2}$) the lasing threshold.

## 2.3   Refractive index sensing and calibration

To measure and quantify the functionalization of our microlasers via their laser spectra, we compare the spectral position of the different lasing modes to an optical model. The model predicts the wavelength associated with each WGM mode using an asymptotic approximation to the Mie scattering problem.[31] In our previous work,[25] we implemented the comparison between experiment and model via a look-up table approach and demonstrated measurements of the refractive index of the medium surrounding the microlasers (referred to as 'external refractive index'); achieving a maximum sensitivity of $5.5 \times 10^{-5}$ refractive index units (RIU). Here, the refractive index sensitivity depends on a combination of spectrometer resolution, signal-to-background ratio, and factors related to the microlasers, e.g. their Q-factor. The high sensitivity of the microlaser sensing is a result of the fact that TM and TE modes have different evanescent field strengths and therefore their sensitivity to changes in external refractive index differs.[31] (For the microlasers used in this study, TE modes typically have a sensitivity to changes in external refractive index of ~20 nm RIU$^{-1}$, while TM modes have a sensitivity of ~30 nm RIU$^{-1}$.)

In addition to the high sensitivity to absolute external refractive index , the approach can also be used to accurately and simultaneously measure the absolute diameter of the microlaser. The latter enables tagging (also referred to as barcoding): An ensemble of microsphere lasers with a statistical variation in size can be coupled to different capture antibodies of interest and by determining the diameter of each microlaser with high accuracy, the lasers and the objects they are coupled to can be tracked and distinguished from each other.

Measuring the absolute external refractive index requires accurate knowledge of the bulk refractive index of the microlasers (referred to as 'internal refractive index' in the following) because analysis of the spectral position of the lasing modes depends on the ratio between the internal and external refractive index, thus requiring at least one of the two to be known in advance.[31] The fluorescent dye that provides optical gain in our microlasers is added via swelling and subsequent trapping of the dye molecules in the PS microspheres.[32] This process could change the refractive index of the microsphere and thus any absolute measurement of external refractive index requires a calibration measurement where the microspheres are measured in a medium with known refractive index. To determine an exact value for the internal refractive index of the microlasers used in this study, lasing spectra were collected and analyzed for 10 microlasers submerged in deionized water, which has a well characterized refractive index.[38,39] We determine an internal refractive index of the microlasers of $1.5924 \pm 0.0004$ which is in good agreement with the literature value for bulk PS ($n_{ps}$ = 1.606 at 490 nm[40]), with the slight reduction attributed to a partial swelling of the microspheres upon dye loading and the longer wavelengths used. Repeating the calibration measurement in aqueous solutions of different concentrations of glucose, i.e. in environments with larger refractive indices, yielded consistent estimates of the internal refractive index (Supporting Information Figure S6). For all external refractive index fits presented in the following, the internal refractive index of the microlaser was set to 1.5924, with microlaser diameter and external refractive index left as free fitting parameters.

## 2.4 Refractive index sensing of antibodies and antigens

The average refractive index close to the surface of our microlasers is expected to increase after each step of the surface functionalization, i.e. it is expected to be lowest for the bare carboxylated microlasers and highest following the formation of the antibody-antigen complex. Therefore, the fitted external refractive index can be used as measure of the local refractive index of the microlaser. To show this, microlaser spectra were characterized at different stages of the functionalization procedure and compared with the unfunctionalized counterparts.

Functionalizing the microlaser with the IgG antibody (α-IgG) results in an increase in the average external refractive index of $5.3*10^{-3} \pm 9*10^{-4}$ RIU compared to unfunctionalized microspheres in phosphate buffered saline (PBS) while the microspheres blocked with casein only showed a minor increase $1.7*10^{-3} \pm 4*10^{-4}$ RIU (Figure 3a). Adding the corresponding antigen, IgG, to antibody-coated and to blocked microsphere lasers resulted in a total increase in average external refractive index, i.e. compared to the bare unfunctionalized microspheres, by $7.1*10^{-3} \pm 7*10^{-4}$ RIU and $2.8*10^{-3} \pm 3*10^{-4}$ RIU, respectively.

To demonstrate the usability of this technique for other proteins, the measurement was repeated with the CRP antibody-antigen complex (Figure 3a). The changes observed have smaller magnitudes but present the same behavior at different functionalization steps.

The evanescent component of the lasing WGMs is largest at the surface where the binding of the antigens takes place and decays exponentially away from the surface. The measured refractive index is therefore the average external refractive index seen by the evanescent component of the analyzed modes and represents a product of the PBS environment and the effect of the bound antibodies and antigens. The larger index increase upon antibody binding (α-IgG and α-CRP) when compared to the antigen binding (IgG and CRP) is thus attributed in part to the larger mode overlap of the antibodies which are located immediately at the microlaser surface [41] and to the increased density of the primary antibody layer when compared to the subsequent antigen layer, accredited to the random immobilization of the antibody which impairs analyte binding.[42] The changes in refractive index are also governed by differences in size and density of the molecular constructs, their binding affinities, binding kinetics and steric effects.

We can estimate the expected relative change in external refractive index ($\Delta n$) due to the binding of proteins on the microlaser surface which can be then linked to the protein concentration within the evanescent volume of the microlasers. Using the surface coverage of ~2.5 mg m$^{-2}$ expected for large adsorbed proteins on PS,[43] we estimated the protein density within the evanescent mode volume. A typical TE mode in our polystyrene microlasers (mode number, $\ell = 99$) has an evanescent field with a 1/e width of ~120 nm (Figure S7). The protein density within the field width is 20 mg ml$^{-1}$ which can be directly linked to refractive index change via the mean protein refractive index increment of 0.190 ml g$^{-1}$.[44] From this calculation, we estimate $\Delta n \sim 4*10^{-3}$ RIU, which agrees with the experimental values obtained for the binding of α-IgG and α-CRP.

In addition to sensing refractive index changes for ensembles of microlasers, we investigated the dynamics of the refractive index change of single microlasers upon antigen binding. As these experiments are not influenced by inherent variability between microlasers (e.g. in size, internal refractive index, antibody coverage, etc.), it is expected that they can monitor the binding of molecules in real time and with drastically improved sensitivity. To allow continuous monitoring, the microlasers were first bound to a glass surface. The antibody functionalized microlasers were coated with biotin functionalized BSA, which binds to the streptavidin on the glass surface. This procedure did not lead to significant changes in the lasing or refractive index characteristics (Figure S8). The WGM spectrum was then monitored continuously while IgG was added to the solution, bringing its final concentration to 12.5 μg ml$^{-1}$ (83 nM). Figure 3b shows the change in external refractive index over time for a representative microlaser as extracted from its WGM spectra, using several improvements as described in the following paragraph. A representative spectrum from this time series is plotted in Figure 3c.

Measuring spectra repeatedly from a single microlaser also allows us to use statistical tools that can significantly improve the refractive index sensitivity as we have previously demonstrated for the analysis of contractility profiles inside live cardiac cells.[25] To achieve this, the slight variability between the fitting results of subsequent spectra can be strongly reduced by also extracting the diameter of the microsphere laser from the recorded spectra. The diameter can then be averaged across all measured spectra and the entire time series is then analyzed again, prescribing this average as a the microlaser diameter and keeping the external refractive index as the only free parameter. Alternatively, if there is a slow drift in effective microlaser diameter, e.g. due to photobleaching, this can be taken into account by fitting the temporal evolution of the diameter with a linear function before re-analyzing the data with a prescribed microsphere diameter and the external refractive index as only free parameter. Furthermore, we eliminate any linear drift in the absolute refractive index by setting the initial slope of the refractive index transient to zero.

After adding the antigen to the edge of the measurement well at $t = 0$ (well dimensions 7.5 x 7.5 mm), there was no significant refractive index change for about 20 minutes while the antigen molecules diffuse toward the microsphere laser. Once they reach the microlaser, the refractive index increases rapidly and then slowly stabilizes after increasing by a total of $\Delta n \sim 1.5*10^{-3}$, similar to changes observed in Figure 3a for α-IgG and IgG binding. The time scale of the dynamical change seen by the laser measurement is in agreement with control measurements conducted with a quartz crystal microbalance (Supporting Information Figure S5).

To illustrate the changes in the WGM spectrum underlying the measured changes in external refractive index, Figure 3d-g shows the individual laser modes at three time points; before the refractive index change, during the attachment of the antigen, and at steady state. All modes red-shift continuously as expected for an increase in the external refractive index. While the magnitude of the mode shifts is comparable to the measured peak width, the shifts can still be clearly resolved via peak fitting / peak localization due to the large signal-to-noise ratio of the microlaser spectra.

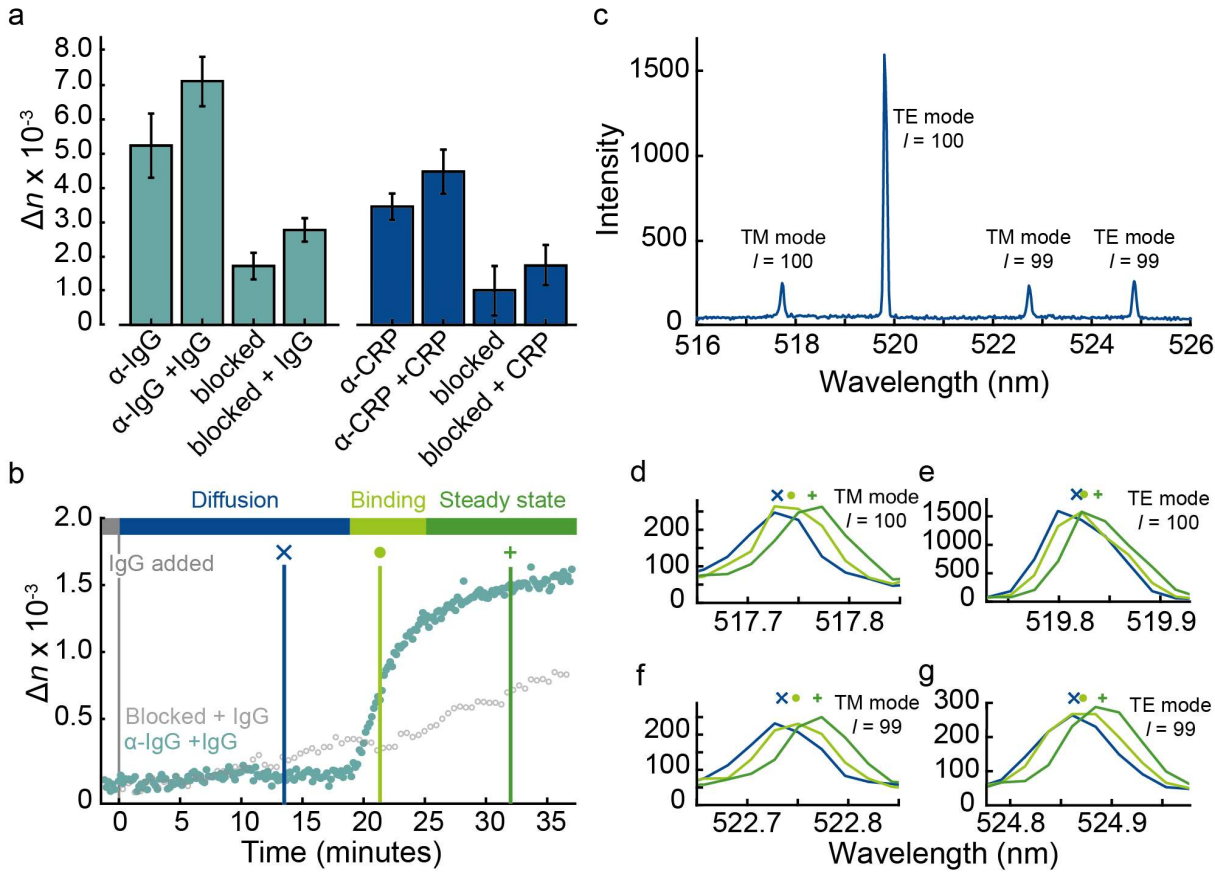

**Figure 3.** Refractive index sensing with WGM microlasers. (a) Change in refractive index at different stages of the microlaser functionalization for two antigens (IgG, aqua green; CRP, dark blue) ($N \geq 10$). The changes are given relative to the external refractive index of bare microlasers in PBS that were neither blocked nor functionalized. (b) Dynamical change of refractive index when IgG is added to an antibody-functionalized microlaser immobilized on a glass surface (aqua green symbols). Also shown is the non-specific binding response of a blocked microlaser (gray). (c) Representative spectrum of the microlaser with the mode numbers and polarization obtained from optical modelling indicated for each peak. (d-g) Spectra of individual modes on a magnified wavelength scale at different time points during transient antibody binding. The symbols above the modes indicate the center positions of each peak as obtained via peak fitting. Symbols and color codes as in (b).

We also evaluated the nonspecific antigen binding dynamics for a blocked microsphere. Here, the refractive index increased linearly with time, which we attribute to binding of the antigen to the microlaser surface as a consequence of incomplete coverage with the blocking agent. Both modalities were performed for different microlasers showing comparable systematic trends and similar absolute refractive index changes, varying only in onset time (Supporting Information Figure S9).

## 2.5 Distribution of absolute refractive indices

Next, we investigated the distribution of the absolute external refractive index for larger sample sets to mimic the situation encountered in high-throughput sensing experiments. Figure 4a depicts the refractive index distribution of bare carboxylated microspheres in PBS, that were neither blocked nor functionalized. Each microlaser spectrum is processed with the fitting algorithm described above and the external refractive index and microlaser diameter were calculated to determine the external refractive index distribution. Fitting this data with a normal distribution gives a maximum refractive index of 1.3354 and $\sigma = 0.0021$ RIU, which is in agreement with the expected value for PBS of 1.3359 (obtained by evaluating the Cauchy coefficients in Ref.[45] for $\lambda = 520$ nm). Similarly, Figure 4b depicts the refractive index distribution of fully IgG functionalized microspheres, i.e. with the IgG antibody and antigen bound, which yields a maximum external refractive index of 1.3405, with an equally narrow distribution ($\sigma = 0.0020$ RIU).

Finally, bare carboxylated microspheres and fully IgG functionalized microspheres were mixed, and the lasing spectra of a random set of microspheres was recorded. Fitting was conducted in the same manner as previously for individual populations, but now using a broader external refractive index range for the fitting (see Experimental section). The resulting distribution and histogram are plotted in Figure 4c. Fitting the distribution of refractive indices with a kernel smoothing function allows for the visualization of the probability density in a multivariant setting and confirms the presence of two populations with refractive index maxima at 1.3361 and 1.3411. Both distribution means are well within the results found for the individual populations and their corresponding standard deviations.

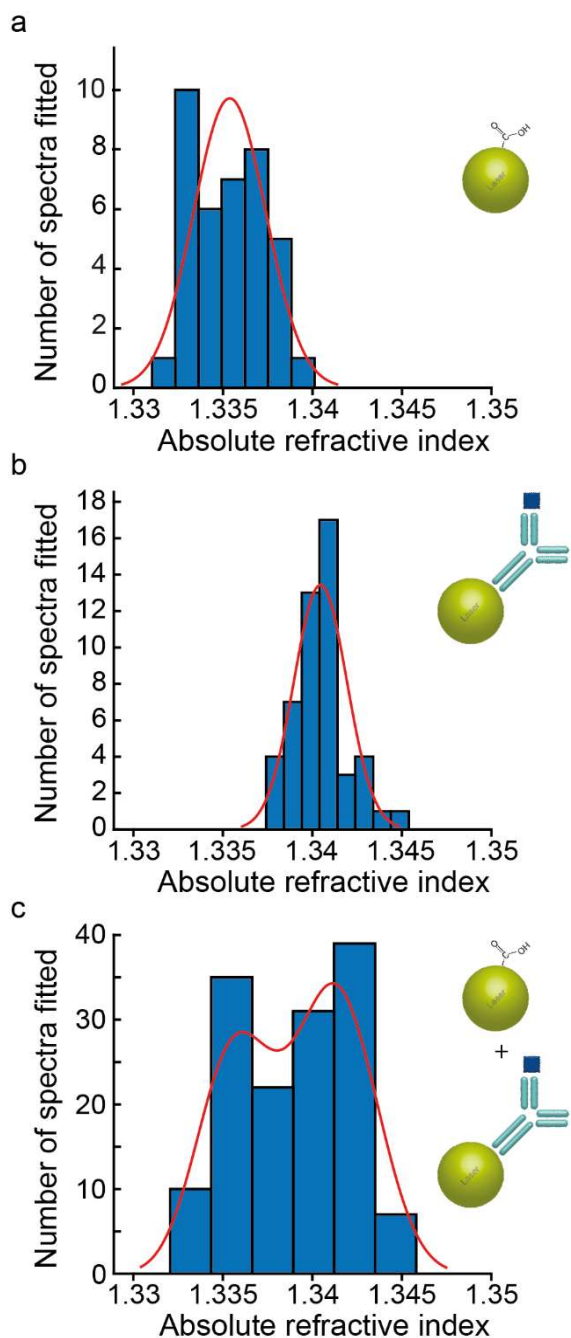

**Figure 4.** Absolute refractive index distribution for different microlaser populations. (a) Histogram of the external refractive index distribution for bare carboxylated microlasers (N = 38) and (b) microlasers functionalized with αIgG and IgG (N = 50). The red lines represent Gaussian fits to the distribution data. (c) Distribution of external refractive index for a mixture of bare carboxylated and αIgG+ IgG functionalized microlasers (N = 144). The red line represents the result of a kernel smoothing function. Histogram bin sizes were determined according to the Freedman-Diaconis rule.

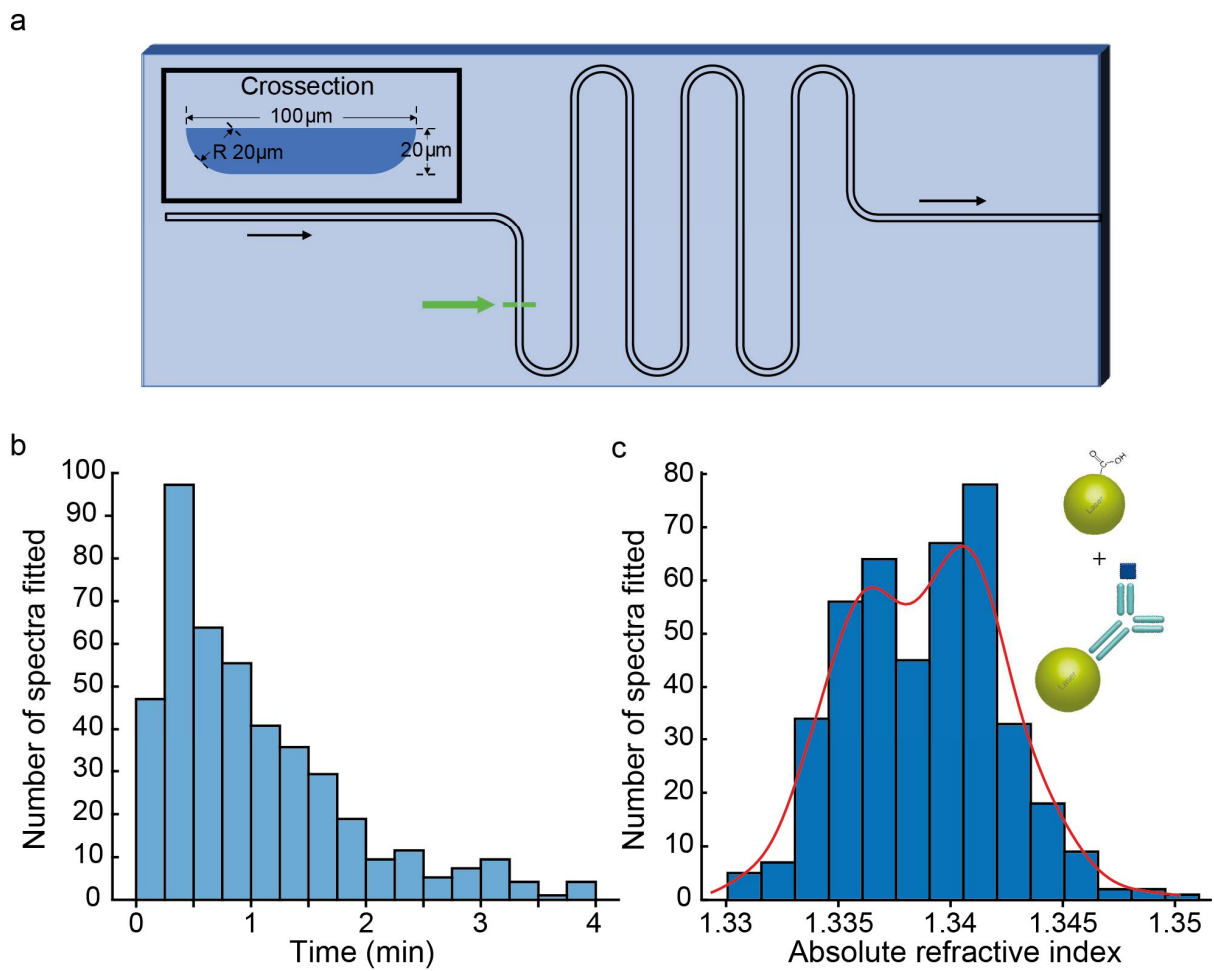

**Figure 5.** High-throughput microfluidic sensing. (a) Schematic of the microfluidic chip. (b) Number of fitted spectra over time. (c) Histogram of refractive indices. The red line represents the result of bimodal gaussian fitting to the distribution data.

Next, to increase the throughput of our measurement, experiments were performed inside a commercially available microfluidic chip (Figure 5a). A blue diode laser set to 1 kHz repetition rate was used as the excitation source, the pump laser beam size was matched to the channel width, and the entire channel width was imaged onto the spectrometer entrance slit. Combined, these conditions ensured the detection of all microlasers passing along the entire width of the channel. Keeping the chip under constant pressure permitted a steady flow of a microlaser dispersion. We used the same microparticle mix as for Figure 4c and optimized the pressure in the microfluidic channel for a flow rate of about 0.03 ml min$^{-1}$, corresponding to 0.2 m s$^{-1}$. This flow speed was sufficiently high to avoid the acquisition of more than one spectrum per microlaser. The rate of lasers measured decayed over time due to uneven loading in this simple setup, possibly due to sedimentation of the microlasers in the input vial (Figure 5b). However, in the first minute of the measurement there are 257 successfully fitted spectra composed of four or more modes. The highest detection frequency was 11 Hz or 93 measured spectra in a 15 s window. The resulting distribution of absolute external refractive index is given in Figure 5c. Due to the large number of microspheres analyzed, the histogram now reveals the bimodal distribution of the two populations more clearly. Normal kernel smoothing uncovers two maxima at refractive indices of 1.3366 and 1.3405, respectively, with a count ratio of 0.88, very similar to the ratio found in the stationary case, 0.82 (Figure 4c). The results demonstrate that precise sensing can be performed with large populations of microsphere lasers and that variations in microsphere size across such populations (Supplementary Information Figure S10) do not prevent measurements of absolute external refractive index. In turn, this enables the direct detection of antibody . In the future, the throughput can be improved further by forcing the particles into a stream passing through the laser beam one by one using microcapillary or acoustic focusing[1], potentially allowing up to 1 kHz acquisition rates even with the equipment used in this study.

## 3    Conclusion and Discussion

In summary, we demonstrate active protein detection with microlasers as small as 10 μm in diameter. Surface functionalization by covalently bound antibodies was used which can be easily adapted to any protein of interest. Specifically, we successfully detect the binding of the biologically relevant antigens, IgG and CRP. The antigens are detected by analyzing the mode position and subsequently calculating the refractive index in the surroundings of the microlasers. Characteristic mode shifts are observed that can be traced directly to the stepwise functionalization protocol. No fluorescently labelled antigens or secondary antibodies are needed to detect the antigen, unlike in conventional immunoassay techniques. Additionally, by monitoring the spectra in real time, the antigen binding dynamics can be observed. Lastly, the refractive index distribution of different functionalized microlasers and mixed populations of microlasers can be determined without prior knowledge of functionalization.

In addition, the spectral signatures of microsphere lasers also allows us to determine the diameter of each microlaser with sub-nanometer precision (Supplementary Information Figure S10).[25,46,47] For the diameter range of 10.8 to 11.4 μm covered by the microsphere lasers used in this study (Supplementary Information Figure S10), and assuming a confidence interval for diameter measurement of 0.5 nm (Supplementary Information Figure S11), one could already in principle distinguish and reidentify up to 1000 microsphere lasers. Using microlaser populations with larger size distributions or by combining multiple populations containing fluorescent dyes emitting in different regions of the spectrum, this number can be extended dramatically. We envision that our microlasers could be used in a flow cytometry setting that is suitable for tagging and sensing in the absence of any fluorescent markers and due to the use of well-established surface chemistry can span to many other antigens and antibodies. The microlasers could also be bound to cells where they could help to evaluate cellular heterogeneity in the context of cellular development and for evaluation of drug therapy at the single cell level.[47,48]

Beyond high throughput screening, functionalized microlasers may be employed to specifically label live cells that present specific antigens with microlasers for optical barcoding. This would be useful to follow the migration of certain cell populations, e.g. circulating tumor cells, in *in vitro* and *in vivo* models. In addition, the functionalized laser platform could be used to detect and monitor specific proteins inside cells.

## 4  Experimental Section

*Surface modification of PS microlasers*: Surfaced modification started from commercially available fluorescent polystyrene beads with exposed carboxyl functional groups (Polyscience Fluoresbrite® YG Carboxylate Microspheres 10.00 μm). An aliquot of beads was dispersed in PBS buffer (10 mM, pH 7.4) at a concentration of 1 mg ml$^{-1}$. The beads were spun down for 5 min at 11000 rpm and resuspended in 1ml of 3.5 mM EDC [(1-ethyl-3-(3-dimethylaminopropyl)carbodiimide hydrochloride), Thermofisher], 5mM Sulfo-NHS (N-hydroxysulfosuccini No-Weigh Format, Thermofisher) in MES buffer (0.1M + 0.5M NaCl, pH 5.6). The solution was incubated at room-temperature (RT) for 15 min and agitated using a bespoke tube-rotator. The sample was then cleaned by spinning down for 5 min at 11.000 rpm and the antibody solution, typically 25 or 50 μg ml$^{-1}$, was added directly. Two antibodies were used, anti-Immunoglobulin G (α-IgG) [Anti-Rabbit IgG, 50 kDa, produced in goat, Sigma Aldrich], and anti-C reactive protein (α-CRP) [26 kDa, Goat Polyclonal, Bethyl Laboratories Inc] was used. Samples were incubated for 2 hours at room temperature (under agitation) and cleaned by spinning down (repeated 3 times). To prevent non-specific protein binding on the biolaser surface, the sample is then left to saturate overnight in a 10% casein (23 kDa) solution in PBS [Casein Blocking Buffer 10x, Sigma Aldrich]. Finally, the sample is cleaned by spinning down and the functionalized beads suspended in PBS buffer. If the sample is not used the following day, it is stored in 1% casein (in PBS). For controls, the blocking solution is added directly after the EDC-NHS and the cleaning step. Immediately prior to the measurement, the appropriate antigen [Human recombinant CRP, 120 kDa, BioVision Inc., or Immunoglobulin G IgG, 150 kDa, Sigma Aldrich) is added (12.5 μg ml$^{-1}$) in PBS and the sample is incubated for 2 hours at room temperature (under agitation) and cleaned by spinning down (repeated 3 times).

*Test-assay for biolaser tagging*: To verify that functionalisation of the beads has been successful, a simple fluorescent assay is run against the appropriate antigen (Human recombinant CRP, BioVision Inc.). To achieve this, a fluorescent tag is covalently bound to the protein using an NHS compound. This is done by adding 6.5 µl NHS-Alexa568 [Alexa Fluor™ 568 NHS Ester (Succinimidyl Ester), ThermoFisher] (15mM in DMSO) to a 1 ml 25 µg ml$^{-1}$ CRP solution (in PBS) and incubating for 1 hour at RT (under agitation). The solution is then filtered using an Amicon Ultra 0.5 mL Centrifugal Filter (Millipore (UK) Limited) with a 30 kDa cut off, spun at 13.4 rpm (12,100 x g) for 10 minutes. In all, 46 µl of supernatant is collected and suspended in PBS (total volume 1 ml). For the assay, fluorescent antigen solution (500 µl) is added to functionalized microlasers (500 µl). The remaining fluorescent antigen solution (500 µl) is added to control microlasers (500 µl). Both solutions are incubated for 2 hours at RT (under agitation). The solutions are cleaned by spinning and suspending in PBS. The supernatant is collected, and the fluorescence measured until the intensity values are below the background, typically after 5-6 cleaning cycles. Finally, fluorescence is measured from the cleaned bead solution. This is done by adding sample and control (70 µl) into a 96-microwell plate and collecting fluorescence from both the microlasers and from the Alexa568 dye (excitation 568 nm, emission 619 nm). Fluorescence from the microlasers themselves is also recorded to confirm their presence (excitation 441 nm, emission 486 nm). Fluorescence emission was recorded for each well using a Synergy H1 microplate reader (Biotek, UK).

*Lasing studies and spectroscopy:* The microlaser solution is diluted to visually identify single spheres. Optical read out is achieved by using custom built inverted microscope setup using a x20 or x40 air objective. Optical pumping is achieved by an OPO laser system (Opotek; pulse duration, 5 ns), tuned to 440 nm. Pump light is focused on the back aperture of the objective yielding a collimated beam with a spot size of 110 µm. Emission from the lasers is collected through the same microscope objective and imaged onto the entrance of a spectrometer coupled to a CCD camera (Andor Shamrock 500i and Andor Newton DU970P- BVF). The spectra are recorded with a high-resolution grating (1200 lines mm$^{-1}$). Neutral density filters are employed to adjust the incident pump fluence.

*Microfluidic studies:* A custom pressure pump system is used to regulate the microlaser flow rate. The system consists of a manual pressure regulator connected to a large pressure vessel with a digital internal pressure gauge, allowing a stable constant pressure. A 1 ml solution of microlasers in PBS is prepared in an Eppendorf tube and placed into the pressure vessel. The solution is fed into the T-junction top connector microfluidic chip (micronit) via 305 µm internal diameter PTFE tubing. To excite the microlasers, a ns-pulsed diode-pumped solid-state laser with emission wavelength 473 nm and repetition rate 1 kHz is used. The laser spot is collimated on the channel and an aperture is used to match the channel diameter. The collection slit of the spectrometer is matched to the cross-section of the channel. The camera is synced to the laser, running at 250 Hz acquisition rate i.e. each frame collects four laser pulses. Data is acquired continuously for 4 minutes.

*Fitting routine:* Peaks in the measured spectra are individually fitted with a Voigt function to accurately find their central position. Two pairs of TM and TE modes are chosen. Initially, a lookup table is generated with a large range of parameters, such as $n_{ext}$ = 1.320 - 1.37, $d$ = 10 - 13 μm for the external refractive index and microsphere diameter, respectively, with step sizes of $\Delta n_{ext} = 5*10^{-4}$ RIU and $\Delta d = 5*10^{-4}$ μm, which determines the number of simulated spectra that need to be generated. The positions of fitted peaks are then compared with the generated spectra, the difference is calculated, the closest match and corresponding parameters are recorded and a smaller parameter space with smaller $\Delta n_{ext}$ ($\Delta n_{ext} = 1*10^{-6}$ RIU) and $\Delta d$ ($\Delta d = 1*10^{-6}$ μm), is generated. Once again, the difference between the fitted and simulated peaks is calculated and if the absolute sum of the difference between simulation and measurements for the four peaks is less than 30 pm, the fit is considered successful.

*Microsphere immobilization for dynamical studies:* Glass bottom petri dishes are cleaned with $O_2$ plasma and filled with a solution of 4% V/V of APTES in IPA which is gently agitated overnight. The petri dish is washed x3 with PBS, incubated with 1 mg ml$^{-1}$ of NHS-biotin for 1h and washed 3x, after which it is incubated with 1 mg ml$^{-1}$ of streptavidin for 1h. The microspheres are prepared as before, but with BSA- biotin instead of casein as a surface blocker, to facilitate biotin-streptavidin binding. To ensure their immobilization, they are left in the dish for 1h with agitation and washed to remove non-attached microlasers.

**Supporting Information**
Supporting Information is available from the from the author.

**Acknowledgements**
This work received financial support from EPSRC (EP/P030017/1), the Humboldt Foundation (Alexander von Humboldt Professorship) and European Union's Horizon 2020 Framework Programme (FP/2014-2020)/ERC grant agreement no. 640012 (ABLASE). M.S. acknowledges funding by the Royal Society (Dorothy Hodgkin Fellowship, DH160102; Research Grant, RGF\R1\180070; Enhancement Award, RGF\EA\180051).

**Supporting Information**

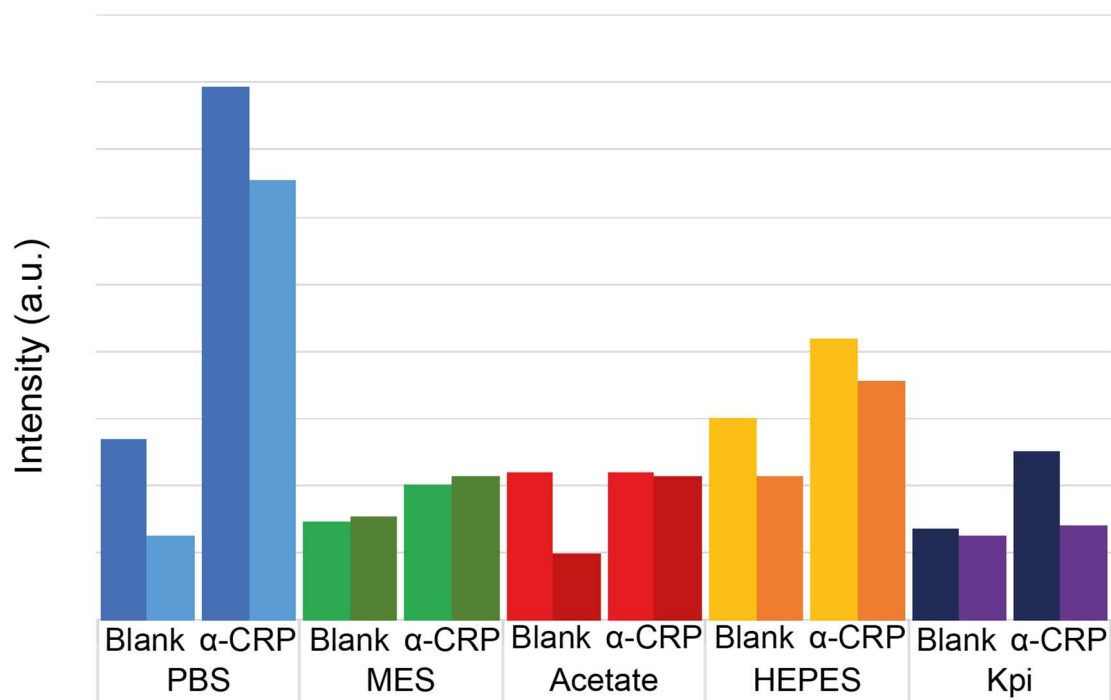

**Figure S1** Optimisation of surface functionalization. Fluorescence intensity of dye tagged CRP antibody (αCRP) in various buffer solutions; phosphate buffered salt (PBS), 2-(N-morpholino)ethanesulfonic acid (MES), acetate buffer, 4-(2-hydroxyethyl)-1-piperazineethanesulfonic acid (HEPES), and phosphate buffer (Kpi). Two repeats were conducted (light and dark colours) for each buffer and antigen (CRP) fluorescence was compared with and without αCRP present on the microlaser surface. For all buffers, a higher fluorescence is obtained for the αCRP sample when compared to the blank counterpart. Note that no blocking of the microlaser was performed for the blank samples to evaluate the degree of unspecific binding. PBS buffer shows the largest change in intensity and therefore reinforces the choice of this buffer to conduct our functionalization assays.

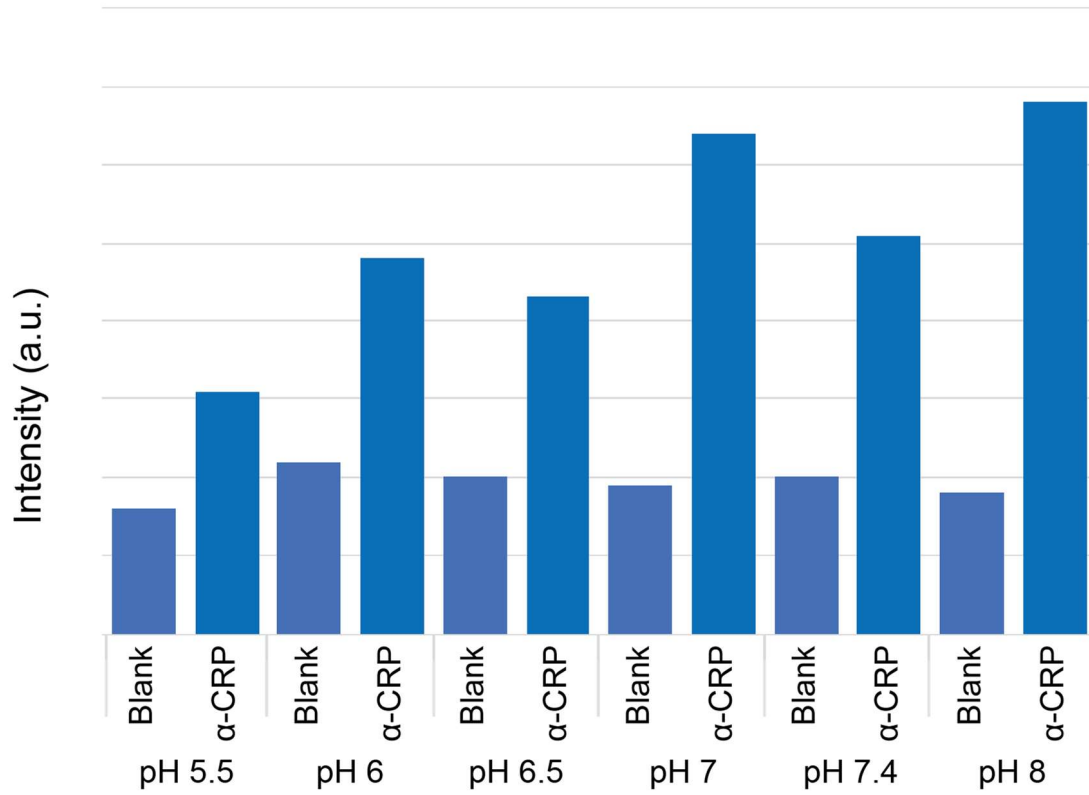

**Figure S2** Fluorescence intensity of dye tagged CRP antibody (αCRP) surface functionalization in PBS buffer solution with different pH values. Values of pH >7 showed the best results, thus for simplicity standard PBS (pH 7.4) is used throughout our assays.

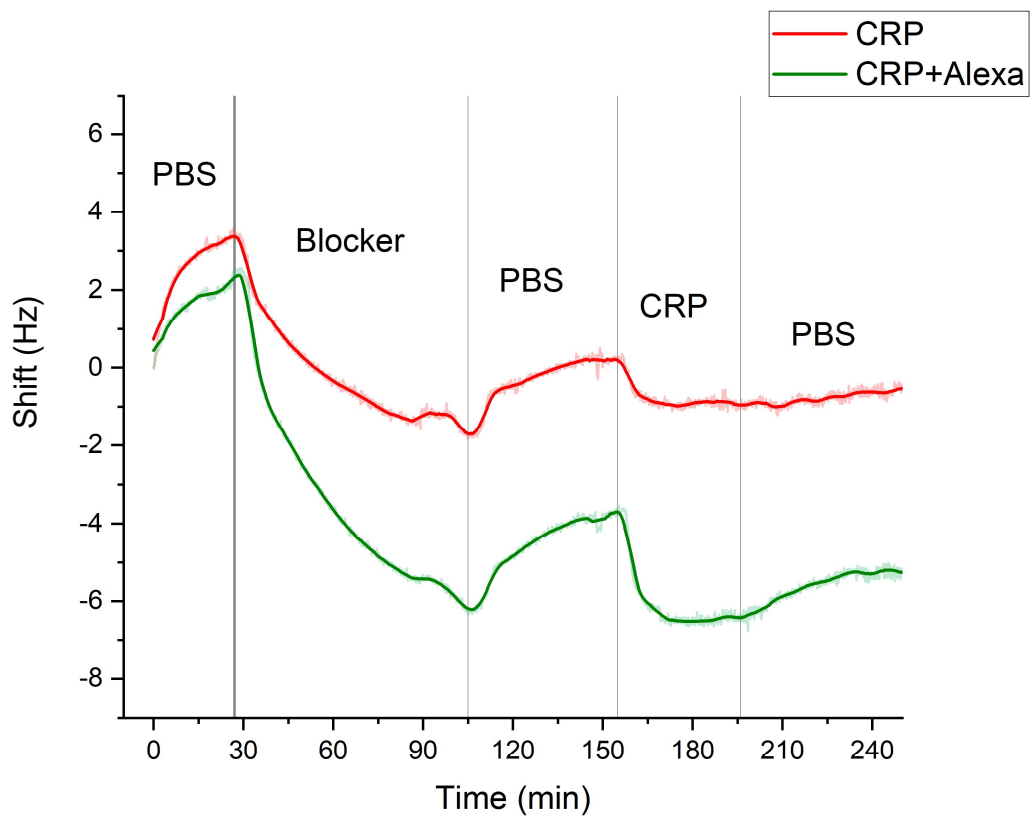

**Figure S3** Verification of maintained antibody-antigen binding upon fluorophore addition (Alexa dye), using a Quartz crystal microbalance (QCM) surface functionalized with αCRP. Both display a pronounced and permanent shift in frequency is observed upon addition of both CRP (red trace) and the CRP-dye complex (green trace), indicating successful interaction with the surface-bound antibody.

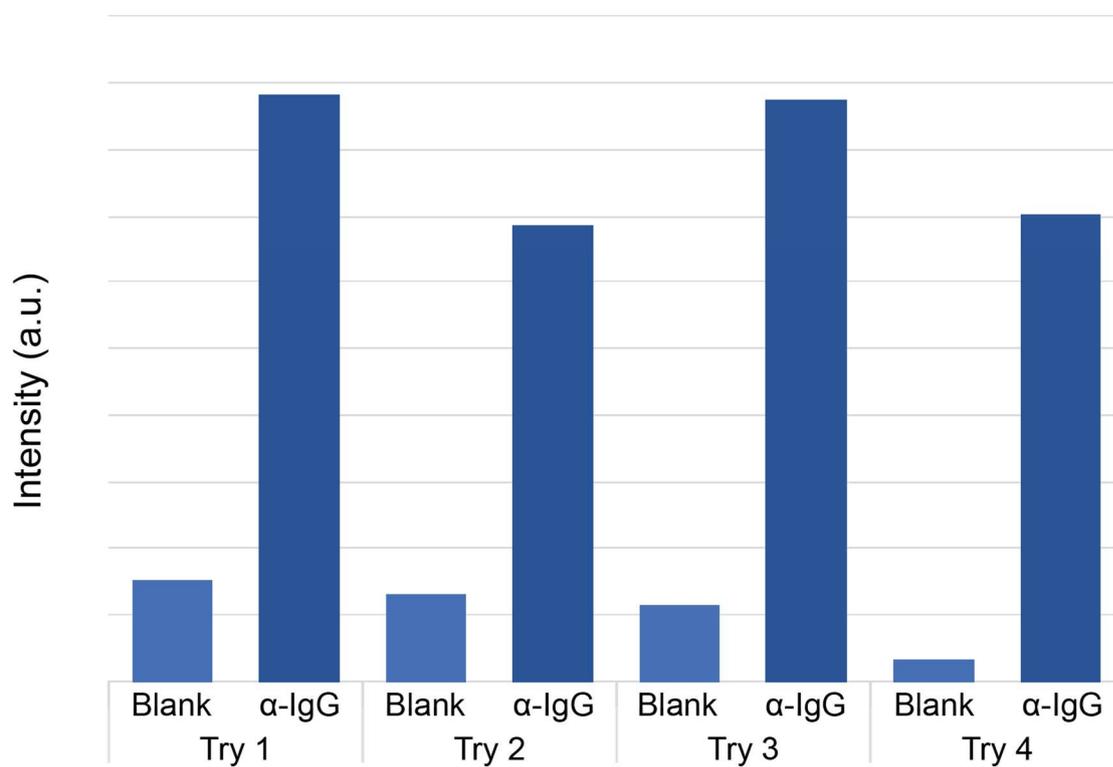

**Figure S4** Luminescence intensity of dye tagged IgG antibody (α-IgG) surface functionalized microlasers. Similar results across all four repeats demonstrate good surface functionalization reproducibility.

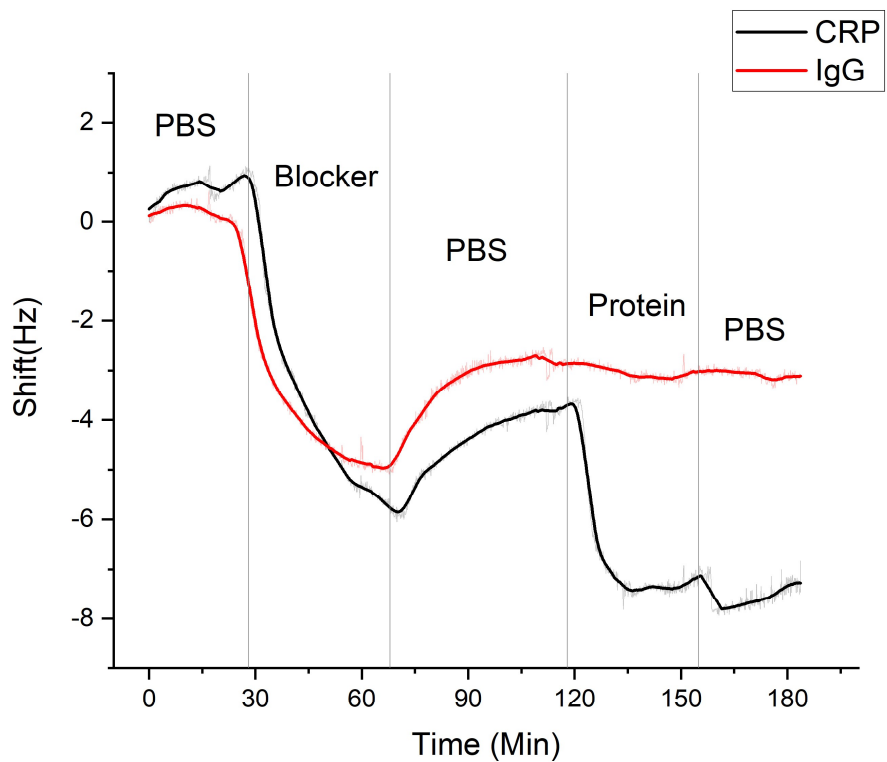

**Figure S5** Verification of specificity of CRP antibody (αCRP) to CRP and not to IgG using QCM. The red trace represents the addition of IgG, demonstrating no significant change in the resonant frequency. The black trace represents the addition of CRP, where a significant frequency shift is observed, indicating successful binding to the antibody.

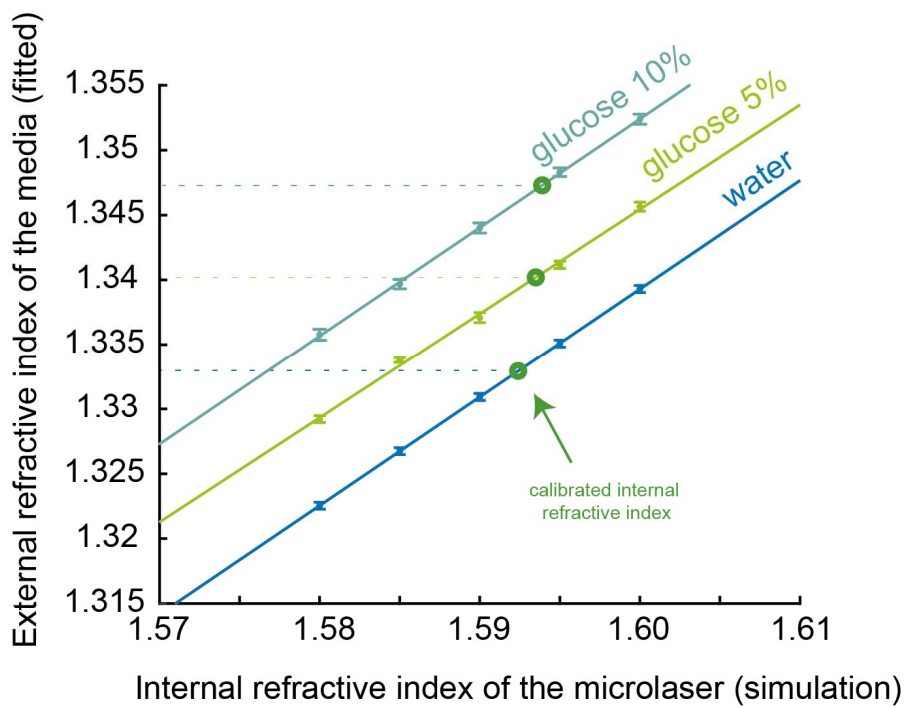

**Figure S6** Calibration curves for the internal refractive index of microlasers. Microlasers (N>10) were submerged in three different liquids: deionised water (dark blue), 5%wt glucose in water (light green) and 10%wt glucose in water (aqua blue). Each spectral set is analysed assuming a different internal refractive index (x-axis), the external refractive index is extracted from the fitted spectra (y-axis) and displayed with the corresponding error bars. Comparing to the literature values of each solution yields an internal refractive index of the microlaser of 1.5924, 1.5935 and 1.5939 for the measurement in deionised water, 5% glucose, and 10% glucose, respectively. The internal refractive index obtained for the microlasers in deionised water is used for all further calculations.

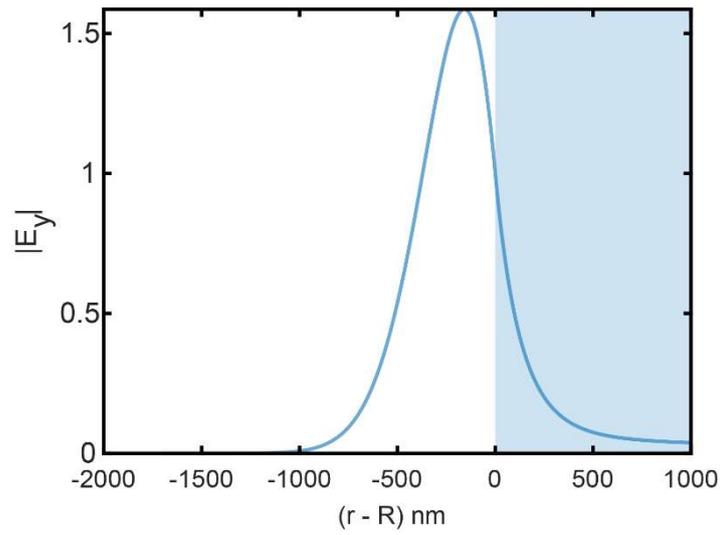

**Figure S7** Simulated electric field crossection for a microlaser (diameter 11 μm; TE mode $m = 99$; $n_{ext} = 1.335$). the white region represents the inside of the microlaser, while the blue region represents the outside of the microlaser. Calculations were performed by means of an open source software [1,2].

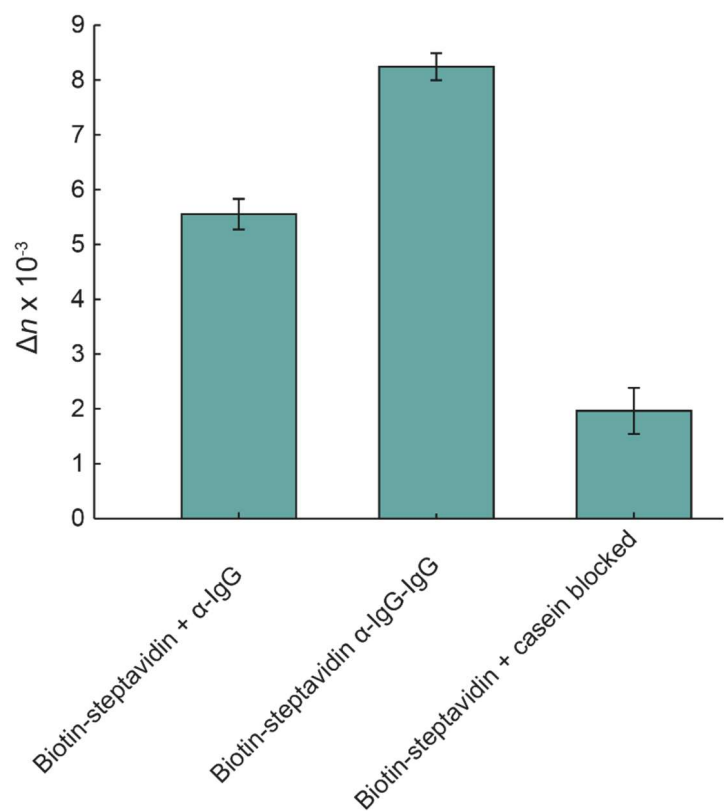

**Figure S8** Refractive index changes induced by microlasers functionalized with biotin groups. No changes in the binding properties of the microlasers were obserbed by the presence of biotin on the microlaser surface.

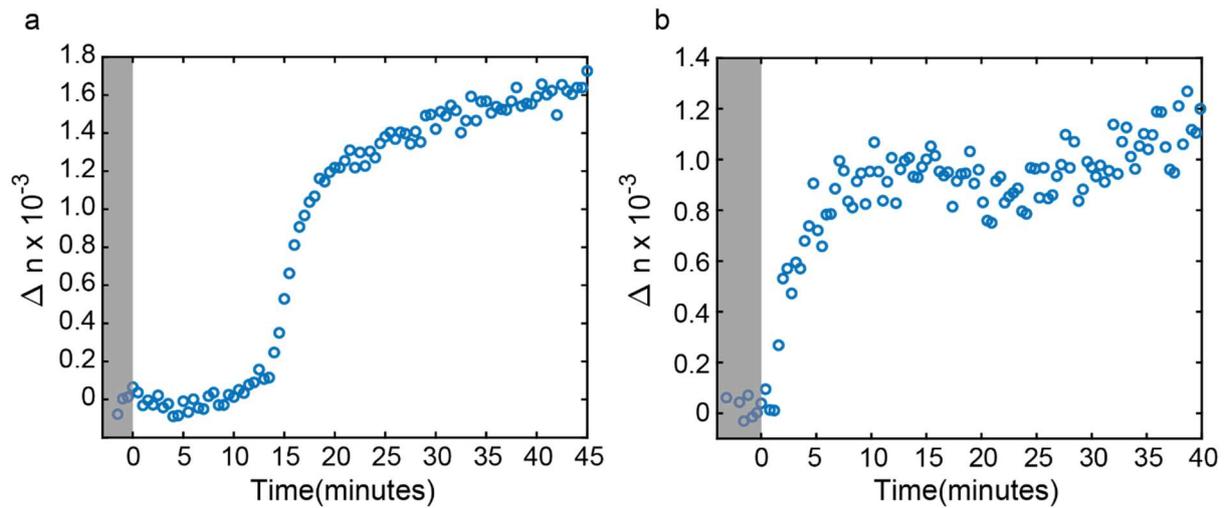

**Figure S9** Additional examples of live protein binding monitoring. (a) IgG is added to the solution for a final concentration of 12.5 ug/ml at $t=0$ minutes. Around $t=10$ minutes the IgG begins binding to the microlaser, stabilising around $t=20$ minutes. (b) IgG is added to the solution for a final concentration of 12.5 ug/ml at $t=0$ minutes. Around $t=1$ minute IgG begins binding to the microlaser, stabilising around $t=10$ minutes.

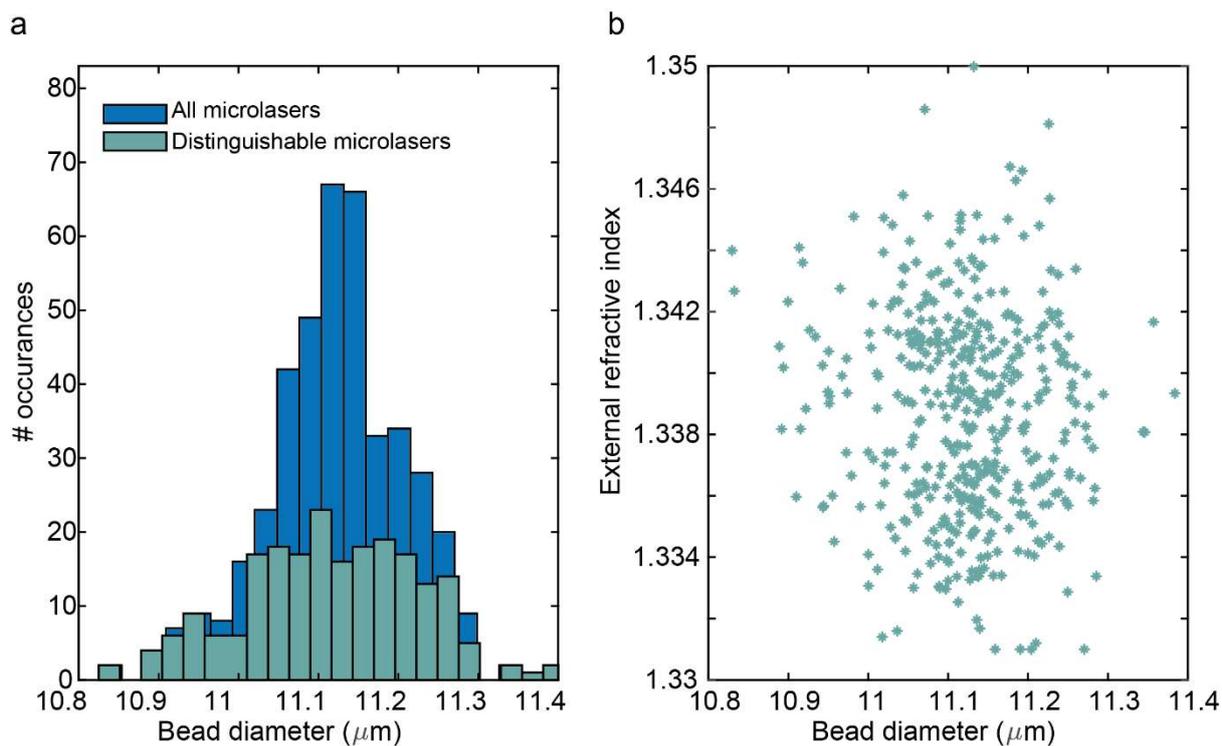

**Figure S10** Microfluidic sensing and spectral barcoding. (a) Histogram of fitted diameters (dark blue) and distinguishable fitted diameters (aqua blue) assuming a minimum separation of 0.5 nm difference. (b) Distribution of microlaser sizes and measured refractive indices for the sample shown in Figure 5.

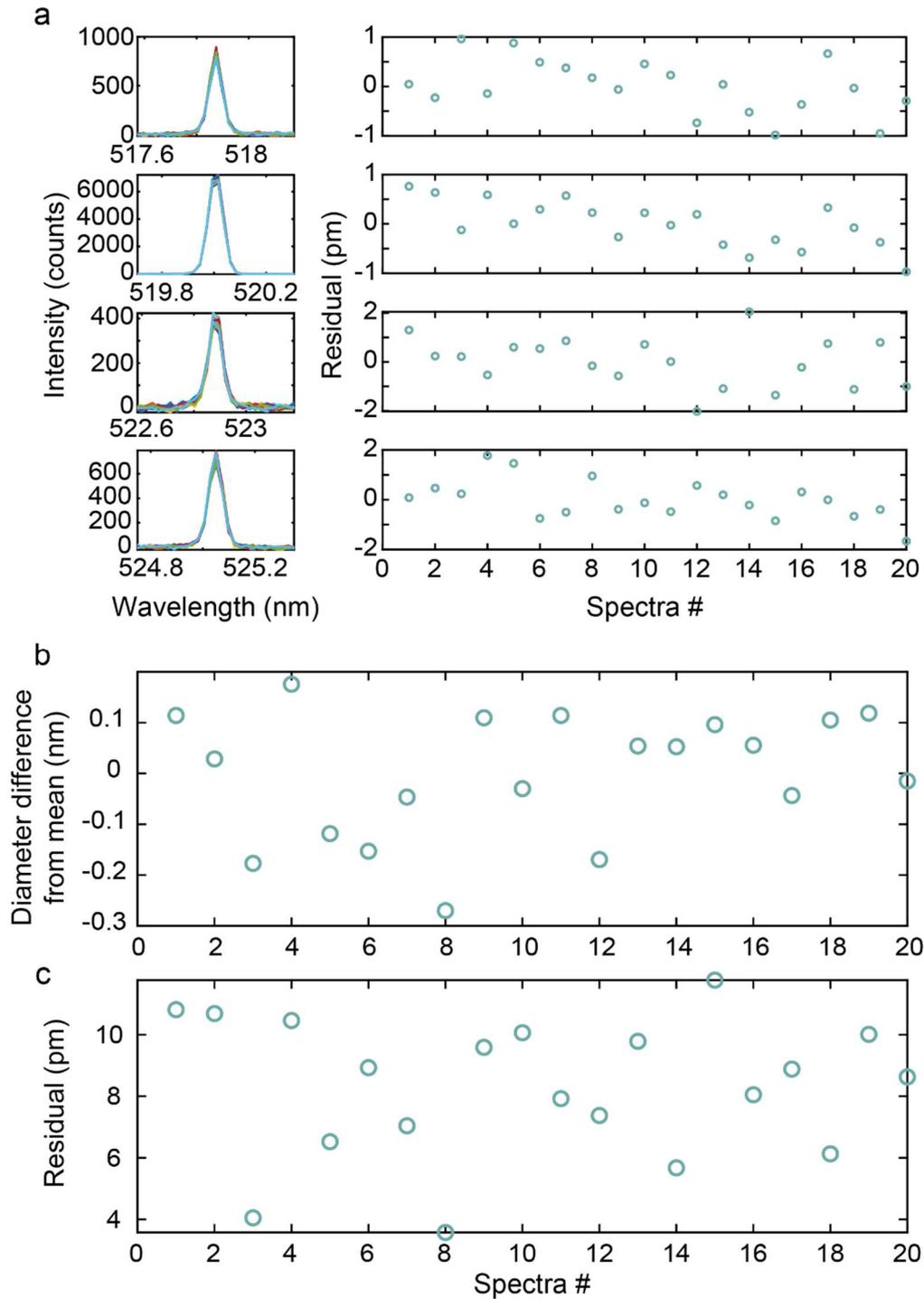

**Figure S11** Error calculations and fit results for a single functionalized microlaser that was measured 20 times. (a) Spectra of 4 modes (left) together with the fitted peak position, displayed as difference from the mean centre position of the mode (right), showing typical differences of ±1pm for high intensity modes and ±2pm for low intensity modes. (b) Difference between the diameter fitted individually for each spectra compared to the mean diameter, obtained from averaging the 20 diameters. The total difference is in the order of 0.5 nm, which sets the boundary condition for the identification of microlasers in Figure S10. (c) Sum of the difference of the numerically calculated mode position and the experimentally obtained (fitted) mode position. We find that this residual varies between 4 pm and 11 pm, showing the excellent agreement of the optical modelling and experimental results.